\RequirePackage[2020-02-02]{latexrelease}
\documentclass[aps,prl,twocolumn,superscriptaddress]{revtex4}
\usepackage{graphicx}
\usepackage{color}
\input epsf
\usepackage{amsmath,amssymb}
\begin{document}
\title{Tunable exchange bias in Y$_3$Fe$_5$O$_{12}$ film on Gd$_3$Ga$_5$O$_{12}$}
\author{Umesh Thuwal}
\affiliation{Department of Physics, Indian Institute of Technology Kanpur, Kanpur 208016, India}
\author{Sumanta Maity}
\affiliation{Department of Physics, Indian Institute of Technology Kanpur, Kanpur 208016, India}
\author{Ruksana Pervin}
\affiliation{Department of Physics, Indian Institute of Technology Kanpur, Kanpur 208016, India}
\author{Rohit Medwal}
\affiliation{Department of Physics, Indian Institute of Technology Kanpur, Kanpur 208016, India}
\author{Joseph Vimal Vas}
\affiliation{The Ernst Ruska-Centre for Microscopy and Spectroscopy, Forschungszentrum Jülich, 52428, Germany}
\author{Yasuhiro Fukuma}
\affiliation{Department of Physics and Information Technology, Faculty of Computer Science and Systems Engineering, Kyushu Institute of Technology, 680-4 Kawazu, Iizuka 820-8502, Japan and Research center for neuromorphic AI hardware, Kitakyushu 808-0916, Japan}
\author{Hervé Courtois}
\affiliation{\mbox{Univ.} Grenoble Alpes, CNRS, Grenoble INP, Institut Néel, Grenoble 38000, France}
\author{Clemens B. Winkelmann}
\affiliation{\mbox{Univ.} Grenoble Alpes, CNRS, Grenoble INP, IRIG-PHELIQS, Grenoble 38000, France}
\author{Anjan Kumar Gupta}
\email{anjankg@iitk.ac.in}
\affiliation{Department of Physics, Indian Institute of Technology Kanpur, Kanpur 208016, India}


\date{\today}
\begin{abstract}
{Ferrimagnetic Y$_3$Fe$_5$O$_{12}$ grown on the (001) surface of paramagnetic Gd$_3$Ga$_5$O$_{12}$ experiences an exchange bias field, which has been attributed to the magnetism of an interface layer between the two materials. We report here that when grown using sputtering and with lower post-annealing temperatures than in previous works, the blocking temperature of the interface magnetic layer is lowered to about 7 K, while still displaying a strong exchange bias. This exchange bias is then found to be tunable between its two extreme values by carefully varying the field cooling protocol. This is attributed to a slow and complex dynamics of the spins of the interface-layer when it is warmed up close to its blocking (or melting) temperature, which is reminiscent of a spin glass.}
\end{abstract}

\maketitle
\section{Introduction}
Achieving precise control over a local magnetization vector is critical for miniaturizing spintronics and magnonics devices. The short-range exchange interaction, arising from overlap of electronic wavefunctions, helps manipulate localized magnetic moments -- a path forward to achieve control over magnetism \cite{wu2010reversible} and spin transport in magnetic materials \cite{vzutic2004spintronics,barman20212021}. This is typically achieved by the pinning of the magnetic moments of a ferromagnet by an antiferromagnetic exchanged bias layer \cite{meiklejohn1956new,nogues1999exchange, kiwi2001exchange,stamps2000mechanisms,malozemoff1987random}. However, realizing a tunable exchange bias suitable for a wide range of applications, such as neuromorphic reservoir computing \cite{NRC-FE1} and quantum magnonics \cite{yuan2022quantum,lachance2019hybrid}, is still elusive.

The record-low Gilbert damping constant and long magnon propagation length of the ferrimagnetic insulator yttrium iron garnet (YIG), \mbox{\emph{i.e.}} Y$_3$Fe$_5$O$_{12}$, \cite{cherepanov1993saga,serga2010yig,hauser2016yttrium} has opened up new avenues for long-distance spin wave transport, information processing and the development of magnon-based spintronic devices. The epitaxially grown YIG thin film on the gadolinium gallium garnet (GGG), \emph{\mbox{i.e.}} Gd$_3$Ga$_5$O$_{12}$, substrates exhibit an interface layer, depending on the pre- and post-growth conditions, which influences the magnetism of the YIG film \cite{popova2003exchange,roos2022magnetization,kumar2021positive,mitra2017interfacial, suturin2018role}. For a film growth, or annealing, at temperatures exceeding 750$^\circ$C, a few nm thick GdIG or GdYIG interface-layer \cite{gomez2018synthetic, mitra2017interfacial, suturin2018role,roos2022magnetization,kumar2021positive} is commonly found. The GdIG interface-layer is ferrimagnetic \cite{roos2022magnetization} and its coupling with YIG provides an exchange bias in YIG which can help control the magnetic properties of YIG thin films.

As a coupled spin system, the YIG/GGG system can also be imagined to display spin frustration, leading to spin-disorder or a spin-glass \cite{cannella1972magnetic} at the
interface. The proximity of a spin-disordered system \cite{ali2007exchange} can enable further control of the magnetic order of the YIG. A spin-glass state exhibits a very wide distribution of relaxation times and energy barriers, leading to effects such as aging and memory that are yet to be fully understood \cite{fisher1988nonequilibrium, dotsenko1993physics,jonason1998memory}.
A spin-disordered state shares certain traits of spin-glass such as multiple metastable spin-states separated by energy barriers and a slow spin dynamics, particularly near and below a certain spin-melting temperature $T_{\rm m}$. Frustration and multiple ground states are also common to many physics and optimization problems, making this topic of wide interest \cite{ferreiro2014frustration,moessner2006geometrical}. Consequently, much ongoing research aims to investigate, better understand and control the distinct ground states of spin-disordered or spin-glass systems.

In this paper, we report on the magnetic behavior of a 40 nm thick and micron-sized epitaxial YIG film on a (001) GGG substrate using a micron-size superconducting quantum interference device ($\mu$-SQUID). Nearly square-shaped magnetization versus applied magnetic field (M-H) loops of this sample, along the easy-axis and at 2.5 K temperature, evidence an exchange bias, whose sign and magnitude depend on the field-cooling history. A systematic history dependence shows the presence of an interface-layer with a frozen spin-disorder which melts at about 7 K temperature. When cooled below that temperature, this layer gets partially spin-ordered under the exchange field of YIG, which, in turn, leads to an exchange bias on YIG. The interface-layer's slow spin-dynamics helps arresting its different spin-ordered states, thus enabling a handle on the YIG exchange bias.

\section{Experimental details}
\begin{figure}[b]
	\centering
	\includegraphics[width=3.4in]{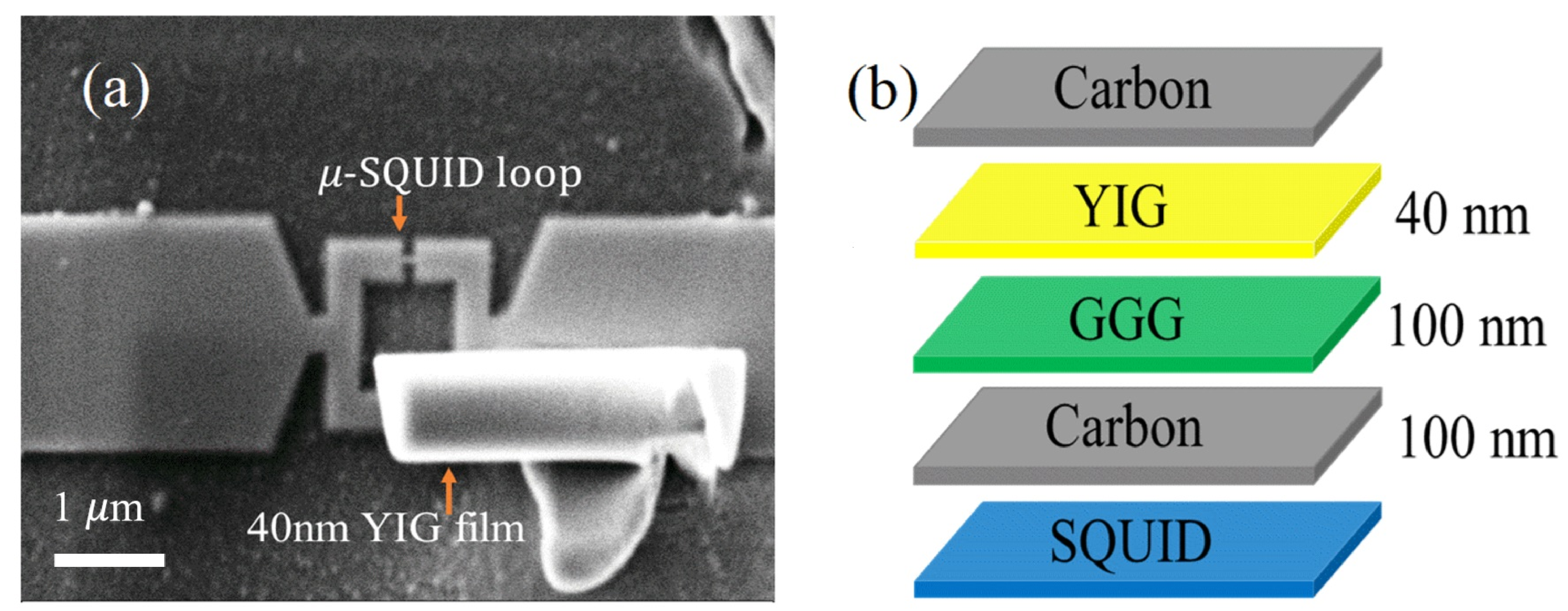}
	\caption{(a) Electron microscopy image of a YIG/GGG sample placed near the Nb $\mu$-SQUID using FIB and nanomanipulation. (b) Schematic illustrating the different layers of the sample on the $\mu$-SQUID.}
	\label{fig:YIGex1}
\end{figure}

A $\mu$-SQUID of Nb with an effective loop area of 1.67 $\mu$m$^2$ is used to probe the magnetic properties of a micron-sized YIG thin film on GGG. The $\mu$-SQUID was fabricated using a 30-nm-thick Nb film on Si substrate using e-beam lithography \cite{paul2020probing}. The 40 nm thick YIG (100) film was grown on a (001) surface of a GGG substrate using rf-sputtering at room temperature followed by annealing at 750$^\circ$C for improved epitaxy and crystallinity \cite{medwal2021facet}. These films have been characterized by X-ray diffraction as well as atomically resolved high angle annular-dark-field scanning transmission electron microscopy (HAADF-STEM). A micron size rectangular piece of this YIG/GGG sample was cut from a mm size substrate using focused ion beam (FIB) milling with gallium ions. After protecting the YIG film by a 100 nm thick carbon layer, the GGG substrate was thinned down to about 100 nm, also with FIB.  This ensures that the overall magnetic signal from the sample is dominated by that of YIG as compared to that of the paramagnetic GGG. Eventually, the $\mu$-SQUID was protected by a 100 nm thick carbon layer before attaching the micron sized YIG/GGG in close vicinity of the $\mu$-SQUID with the help of a nano-manipulator. Figure \ref{fig:YIGex1}(a) shows an electron microscope image of the actual studied sample while \mbox{Fig.} \ref{fig:YIGex1}(b) depicts its different layers.

The $\mu$-SQUID is operated in non-hysteretic mode and with a low temperature commercial SQUID-array amplifier in a 1.3 K base temperature liquid helium cryostat with a 3D vector magnet \cite{paul2020probing}. The sample holder also has a heater and a temperature sensor. A temperature controller is used to control the sample temperature with better than 10 mK temperature stability over several hours and up to 20 K temperature. All the magnetic measurements were performed at 2.5 K temperature while the field cooling was carried out from up to 20 K temperatures. The measurements were done using a magnetic field in the $\mu$-SQUID plane, which is parallel to the YIG film plane, and along the magnetic easy direction of the YIG film \cite{suppl-info}.

\section{Nature of the Interface}
Gomez-Perez \mbox{\emph{et al.}} \cite{gomez2018synthetic} reported a structural and compositional analysis of the interface of a YIG/GGG sample using HAADF-STEM and energy dispersive X-ray (EDX) mapping. They found a relative shift \cite{gomez2018synthetic} of about 2-3 nm between Y and Fe fronts, indicating the presence of Fe beyond Y, and a similar shift of about 2 nm between Gd and Ga fronts. They concluded the presence of a 2-3 nm thick GdYIG, \mbox{\emph{i.e.}} (Y$_{1-x}$Gd$_x$)$_3$Fe$_5$O$_{12}$, interlayer at the YIG/GGG interface which behaves as a compensated ferrimagnet \cite{roos2022magnetization}. This interlayer appears in YIG thin films on GGG with a processing temperature above 750$^\circ$C \cite{gomez2018synthetic,roos2022magnetization,kumar2021positive,mitra2017interfacial} and it explains the observed exchange bias in YIG \cite{roos2022magnetization,kumar2021positive,mitra2017interfacial} at low temperatures \cite{roos2022magnetization} or at room temperature \cite{kumar2021positive}. There are also reports of a non-magnetic dead layer at the YIG/GGG interface \cite{suturin2018role}. Thus the magnetic properties of this interface are very sensitive to the growth conditions and the orientation of the GGG substrate \cite{medwal2021facet}.

\begin{figure}[b]
	\centering
	\includegraphics[width=3.4in]{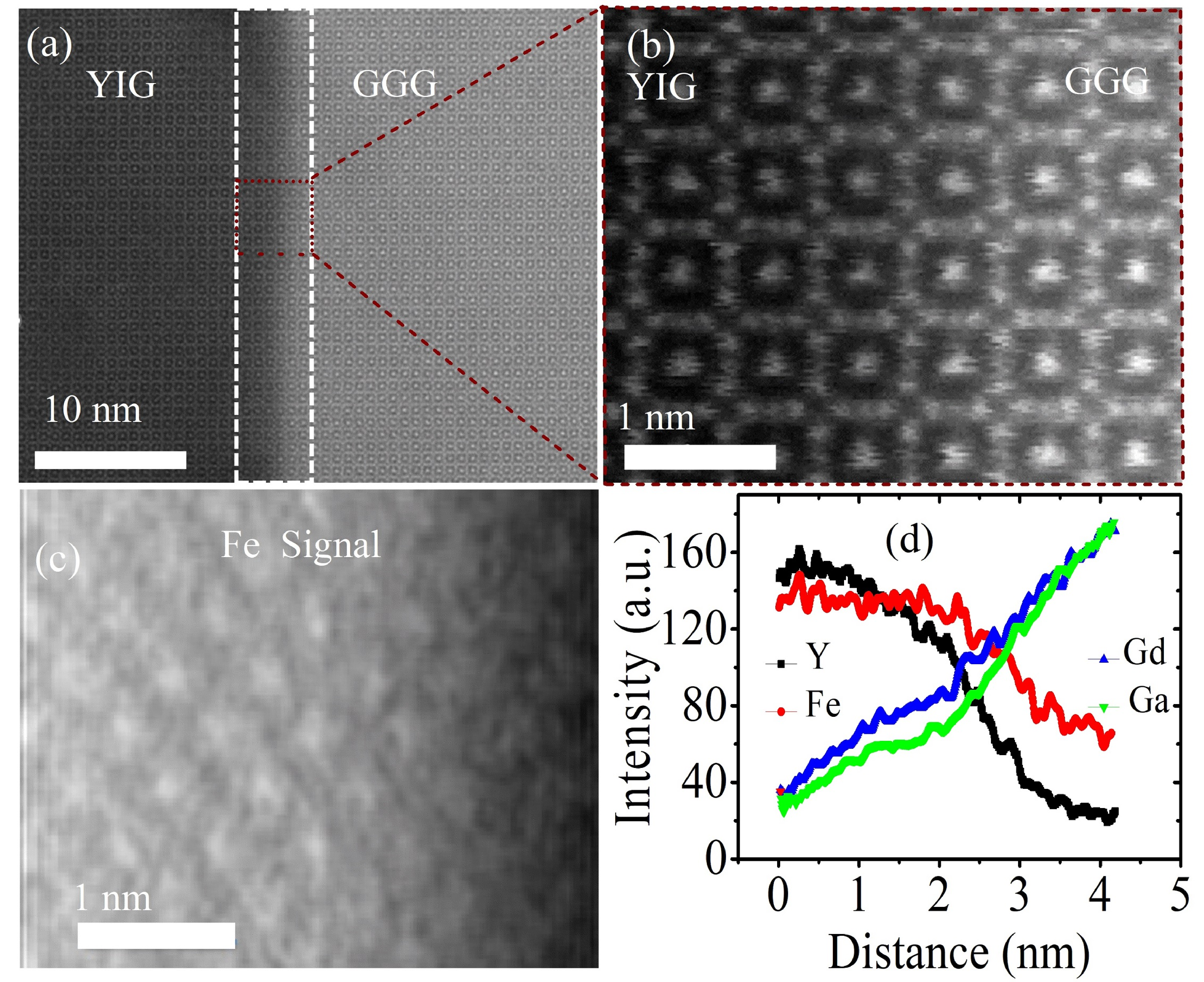}
	\caption{(a,b) HAADF-STEM images of the interface region of the YIG film on GGG (100) substrate showing the atomic lattice with a gradual contrast change at the interface. (c) Fe-EDX map acquired simultaneously with the image in (b). The other elements maps are presented in the suppl-info \cite{suppl-info}. (d) Plot of the average of the line profiles, perpendicular to the interface, to show the variation of Y, Fe, Gd, and Ga EDX signals' across the interface.}
	\label{fig:YIGex2}
\end{figure}

Figure \ref{fig:YIGex2}(a,b) show HAADF-STEM measurements on our studied YIG thin film. The atomic resolution images show a defect-free and atomically matched interface with a continuous variation in contrast across the interface. Figure \ref{fig:YIGex2}(c) shows an EDX image of Fe with a diffused contrast over a scale of a few nanometers, slightly above the resolution (1 nm) of this elemental mapping technique. Figure \ref{fig:YIGex2}(d) shows the profile of Y, Fe, Gd and Ga EDX signals across the interface. There is a relative shift of about 1 nm between Fe and Y fronts, which is smaller than that observed by Gomez-Perez \mbox{\emph{et al.}} \cite{gomez2018synthetic}. On the other hand, we do not see a significant shift between Gd and Ga fronts although they appear to be more diffused than Fe and Y. From these comparisons, we conclude that the YIG/GGG interface here has a different character than that of Gomez-Perez \mbox{\emph{et al.}} \cite{gomez2018synthetic} or others \cite{roos2022magnetization,kumar2021positive,mitra2017interfacial,suturin2018role} where the film growth involved temperatures above 750$^\circ$C. From the Fe front progressing by about 1 nm ahead of Y, some Fe diffusion into GGG can be inferred while there is negligible selective diffusion of Ga over Gd into YIG.

The atomic magnetic moments in a unit cell of YIG arise from 24 Fe$^{3+}$ ions at tetrahedral A-sites and 16 Fe$^{3+}$ ions at octahedral B-sites while the octahedral C-sites are occupied by non-magnetic Y$^{3+}$ ions \cite{musa2017structural}. A-A and B-B site interactions are strongly ferromagnetic but the A-B interaction is antiferromagnetic due to super-exchange through oxygen, which leads to the ferrimagnetic behavior of YIG. GGG is a spin-frustrated system with the magnetic moments arising from Gd ions that occupy two interpenetrating, corner-sharing triangular sublattices, resulting into geometric frustration \cite{schiffer1995frustration}. No magnetic long-range order is observed above 0.4 K in GGG. The diffusion of Fe into GGG due to annealing can lead to Fe-Gd, Gd-Gd and Fe-Fe linkages. The resulting magnetic interactions can have either sign depending on local bond distances and angles \cite{gorbatov2021magnetic}. Further, the geometric frustration arising from the triangular sublattices of the garnet can easily lead to a spin-disorder in this intermixed YIG-GGG interface.

\begin{figure}[bt!]
	\centering
	\includegraphics[width=3.4in]{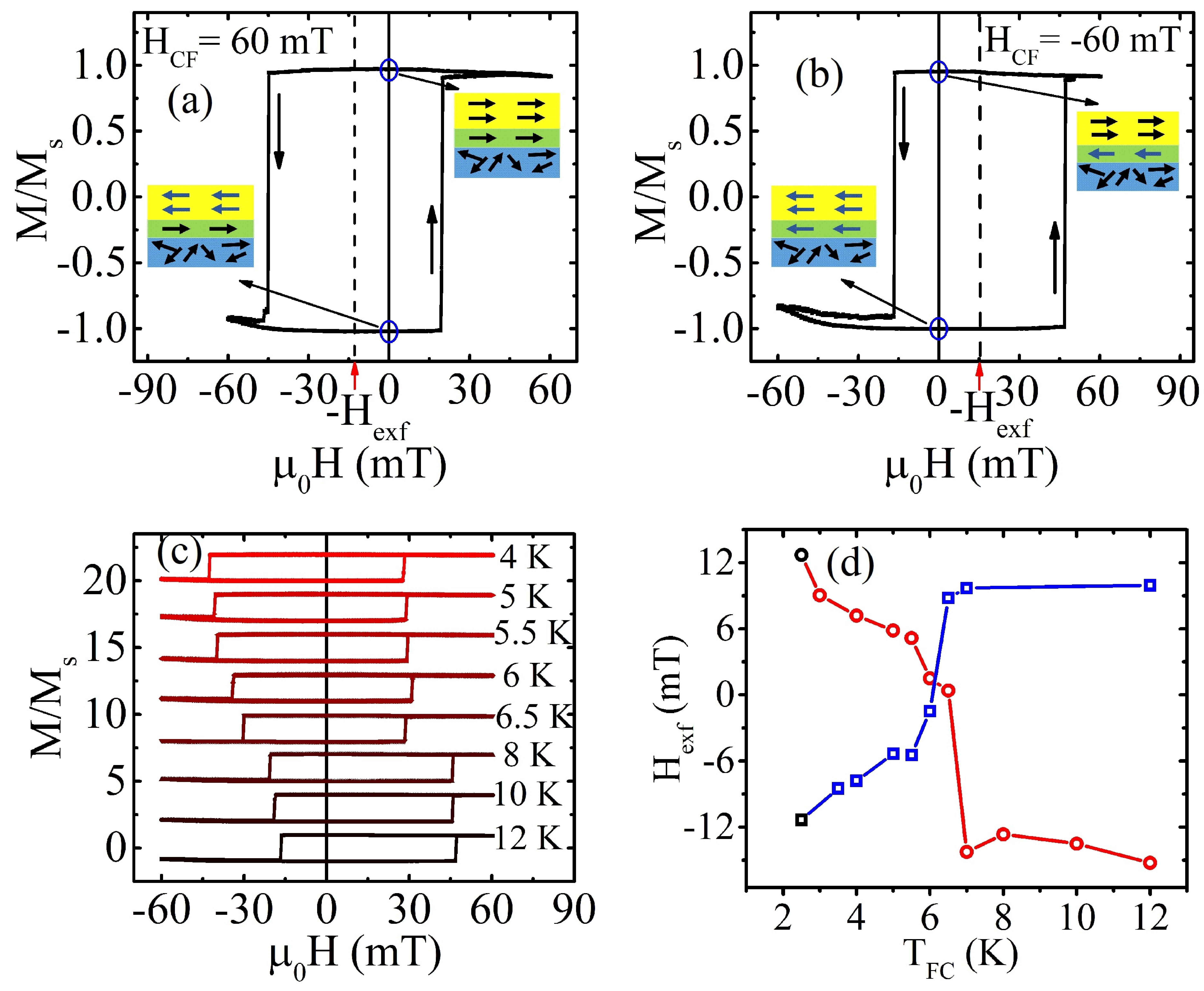}
	\caption{(a,b) M-H loops of YIG measured at 2.5 K after field-cooling the sample from temperature $T_{\rm FC}=T_{\rm RST}=$ 12 K and under cooling fields $H_{\rm CF}=$ +60 and -60 mT, respectively. The vertical dashed lines mark the center of the respective loops and represent the exchange bias, \mbox{\emph{i.e.}} $-H_{\rm exf}$. The schematics in (a,b) insets depict the spin-states of the YIG (yellow), the interface (green) and the GGG (blue) layers at the marked points of the M-H loops, assuming a ferromagnetic exchange coupling between YIG and the inter-layer. (c) M-H loops (with vertical offsets, for clarity) measured after different field cooling protocols with each involving a reset at +60 mT followed by field-cooling at -60 mT from different field-cooling temperature $T_{\rm FC}$. (d) The red plot shows the variation of the exchange field $H_{\rm exf}$ with $T_{\rm FC}$ as deduced from the M-H loops of (c). The blue plot shows the $T_{\rm FC}$ dependence of $H_{\rm exf}$ for field cooling protocols with opposite field direction \cite{note-diff-runs}. Thus, it involved a reset at -60 mT followed by field-cooling at +60 mT from different field-cooling temperature $T_{\rm FC}$.}
	\label{fig:YIGex3}
\end{figure}

\section{Exchange bias at 2.5 K}
Figure \ref{fig:YIGex3}(a,b) show the nearly box shaped M-H loops, measured at 2.5 K temperature, for a magnetic field $H$ applied along the easy axis and after field cooling with two opposite cooling-fields $H_{\rm CF}= \pm60$ mT, from a field-cooling temperature $T_{\rm FC}=12$ K. The M-H loops exhibit a horizontal shift of -12.7 mT and +15.2 mT, amounting to an exchange bias \cite{note-diff-runs}. The exchange bias is common in ferromagnetic-antiferromagnetic (FM-AFM) systems \cite{nogues1999exchange} after the field-cooling through the ordering temperature of the AFM above which it is paramagnetic. This enables AFM surface spins alignment with the FM that exert an exchange field on FM. In our system, the Curie temperature of the ferrimagnetic YIG is about 560 K while GGG is paramagnetic down to 0.4 K \cite{petrenko1999magnetic}.

As already reported in earlier studies \cite{roos2022magnetization,kumar2021positive}, we attribute the observed exchange bias to an interface layer at YIG/GGG interface.
The interface-layer, when cooled through its spin-ordering or spin-melting temperature $T_{\rm m}$, freezes into a certain spin-state. As discussed later, the intrinsic low temperature state of this interface layer seems to be spin-disordered, like that of a spin-glass, but it acquires a spin-order under YIG's exchange field. With $T_{\rm m}$ being much smaller than YIG's Curie temperature, the YIG's exchange field on the interface-layer is unavoidable. The schematics in \mbox{Fig.} \ref{fig:YIGex3}(a,b) insets depict the spin-states of the GGG, the interface-layer and the YIG at two different points of the M-H loops. While our measurements cannot deduce the sign of exchange interaction between YIG and the interface-layer, this schematic uses a ferromagnetic exchange coupling for illustration only.

\begin{figure*}[hbt!]
	\centering
	\includegraphics[width=6.8in]{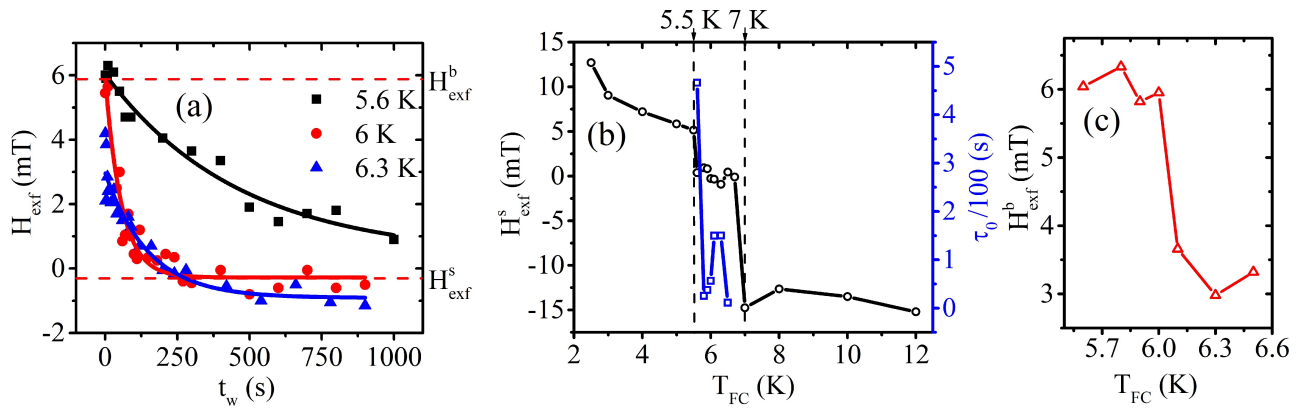}
	\caption{(a) Variation of the exchange field $H_{\rm exf}$ with the waiting time $t_{\rm w}$ for three field-cooling temperature $T_{\rm FC}$ values together with corresponding fits (continuous lines) to \mbox{Eq.} \ref{eq:FSS}. These $H_{\rm exf}$ have been deduced from the M-H loops, some of which are presented in the suppl-info \cite{suppl-info}. The discontinuous red lines mark the fitted $H^{\rm b}_{\rm exf}$ and $H^{\rm s}_{\rm exf}$ for the 6 K plot. (b) Variation of $H^{\rm s}_{\rm exf}$ (black) and $\tau_0$ (blue) with T$_{\rm FC}$. Note the non-monotonic diminishing of $\tau_0$ starting from a large value at 5.6 K to a negligible value at 7.0 K with a bump in between. (c) Variation of $H^{\rm b}_{\rm exf}$ with the field-cooling temperature T$_{\rm FC}$. }
	\label{fig:YIGex4}
\end{figure*}

For temperatures well below $T_{\rm m}$, the partially spin-ordered interface-layer exerts an exchange field $H_{\rm exf}$ on YIG. Together with an external field $H$, the total field experienced by YIG becomes $H_{\rm tot}=H+H_{\rm exf}$. Thus, for a given coercive field $H_{\rm C}$ of YIG, the magnetic reversals are observed at $H_{\rm tot}=\pm H_{\rm C}$ or at applied field $H=\pm H_{\rm C}-H_{\rm exf}$. The exchange field $H_{\rm exf}$ has the same sign as the cooling-field $H_{\rm CF}$ and it is the opposite of the exchange bias defined as the M-H loop shift. Thus, for a positive $H_{\rm CF}$ the M-H loop shifts towards negative applied fields.

The exchange field $H_{\rm exf}$ can be taken as a measure of the interface layer's average spin-order. This spin order is influenced by the exchange-field of YIG and the temperature. The YIG's exchange field will drive the interface-spins to a suitable minimum energy state provided the temperature is high enough to allow the thermal activation out of local energy minima. The disordered interaction between the interface-layer's spins can lead to spin frustration and trapping in one of the numerous minima with or without some spin-order. This may enable a robust $H_{\rm exf}$ at low temperatures. For higher temperatures $T>T_{\rm m}$, the interface layer's spins will respond immediately to a change in the direction of YIG's exchange field. The latter is determined by the direction of its magnetization. Note that the exchange bias is generally a small fraction of the actual inter-spin interaction \cite{kiwi2001exchange}. Thus the YIG's exchange field is expected to dominate over the applied magnetic field. The interface layer's spins, particularly in case of a disordered state, may slowly relax, even below $T_{\rm m}$, leading to changes in $H_{\rm exf}$. Spin relaxations with extremely wide distribution in relaxation-times and other related effects, such as aging and memory, are well known in spin-glasses \cite{vincent2022spin}.

In order to determine the melting temperature $T_{\rm m}$ of the interface-layer spins, the M-H loops at 2.5 K temperature were acquired after cooling from different field-cooling temperatures $T_{\rm FC}$. For each $T_{\rm FC}$, the sample was first field-cooled under a +60 mT applied field from a fixed reset temperature $T_{\rm RST}=12$ K down to 2.5 K to reset any previous history and to obtain $H_{\rm exf}\approx 12.7$ mT. As seen later, this $T_{\rm RST}$ is nearly double of $T_{\rm m}$. The sample is then warmed-up to $T_{\rm FC}$ followed by the application of an opposite -60 mT field. After a waiting time $t_{\rm w}=60$ s, the temperature was brought down to 2.5 K, still under a -60 mT field. The field cooling protocol's flow-chart can be found in the suppl-info \cite{suppl-info}. It takes few seconds to cool from 12 to 2.5 K. The M-H loops are then measured for different $T_{\rm FC}$ values. Figure \ref{fig:YIGex3}(c) shows the third measured M-H loops exhibiting a $T_{\rm FC}$-dependent shift. The M-H loops also show some training effects in first few loops, which is discussed later.

Figure \ref{fig:YIGex3}(d) shows a combination of two $H_{\rm exf}-T_{\rm FC}$ plots where the gradual melting was probed using two reset-states at $\pm$60 mT. Considering the +60 mT reset field data (red curve), a gradual decline in $H_{\rm exf}$ value is seen with rising $T_{\rm FC}$ from 2.5 K and to nearly a zero value. This is followed by an abrupt jump near 7 K beyond which $H_{\rm exf}$ takes a nearly fixed value close to -15 mT. From this we conclude a $T_{\rm m}\approx 7.0$ K. The -60 mT reset field data show a nearly symmetric behavior \cite{note-diff-runs}. Overall, a remarkable control on $H_{\rm exf}$ value between +12.7 and -15.2 mT is demonstrated.

\section{Gradual melting and arrest of interface spin-disorder}
The dependence of the exchange field $H_{\rm exf}$ on the waiting time $t_{\rm w}$ for different $T_{\rm FC}$ values below $T_{\rm m}$ is probed next. This was done only for the positive reset field. The experimental procedure is the same as the one used to produce \mbox{Fig.} \ref{fig:YIGex3}(c,d) data, but with a varying waiting time $t_{\rm w}$ at the temperature $T_{FC}$. Note that we cannot probe $t_{\rm w}<5$ s in our setup as this is the typical cooling time from 12 to 2.5 K. About three to five M-H loops were then measured at 2.5 K after this -60 mT field-cooling with different $T_{\rm FC}$ and $t_{\rm w}$ values.

Figure \ref{fig:YIGex4}(a) shows the dependence of $H_{\rm exf}$ on $t_{\rm w}$ as deduced from the M-H loop shifts. Some of these M-H loops are presented in the suppl-info \cite{suppl-info}. We observe a sharp change in $H_{\rm exf}$, happening at times $t_{\rm w}<5$ s from $H_{\rm exf}=12.7$ mT and not shown in \mbox{Fig.} \ref{fig:YIGex4}(a), followed by a slow relaxation. We denote the exchange field value in the zero waiting time limit by $H_{\rm exf}^{\rm b}$ which is $T_{\rm FC}$-dependent. When the waiting time increases, $H_{\rm exf}$ slowly relaxes to a saturation value $H_{\rm exf}^{\rm s}$ with a characteristic time $\tau_0$. Thus, an exponential fit is made using the relation,
\begin{equation}
H_{\rm exf} = H_{\rm exf}^{\rm s} - (H_{\rm exf}^{\rm s}-H_{\rm exf}^{\rm b}) \exp(-t_{\rm w}/\tau_{\rm 0}).
\label{eq:FSS}
\end{equation}
These fits are shown in \mbox{Fig.} \ref{fig:YIGex4}(a) and the variation of the fitting parameters, \mbox{\emph{i.e.}} $H_{\rm exf}^{\rm s}$, $H_{\rm exf}^{\rm b}$ and $\tau_{\rm 0}$, with $T_{\rm FC}$ is summarized in the plots in \mbox{Fig.} \ref{fig:YIGex4}(b,c). Here $H_{\rm exf}^{\rm s}$ has been plotted over a broad temperature range including the
the regimes where no slow relaxation is observed. Thus, for $T_{\rm FC}<5.5$ K, no slow relaxation is found while for $T_{\rm FC}>7.0$ K, a steady state value $H_{\rm exf}\sim -15$ mT is found. From this we conclude a slow spin-melting component over a 5.5 to 7.0 K temperature range where $H_{\rm exf}$ changes from an initial value $H_{\rm exf}^{\rm b}$ to near zero over a time scale $\tau_0$. Eventually, a fast melting to a paramagnetic state of the interface-spins occurs at 7.0 K.

\begin{figure}[hbt!]
	\centering
	\includegraphics[width=3.1 in]{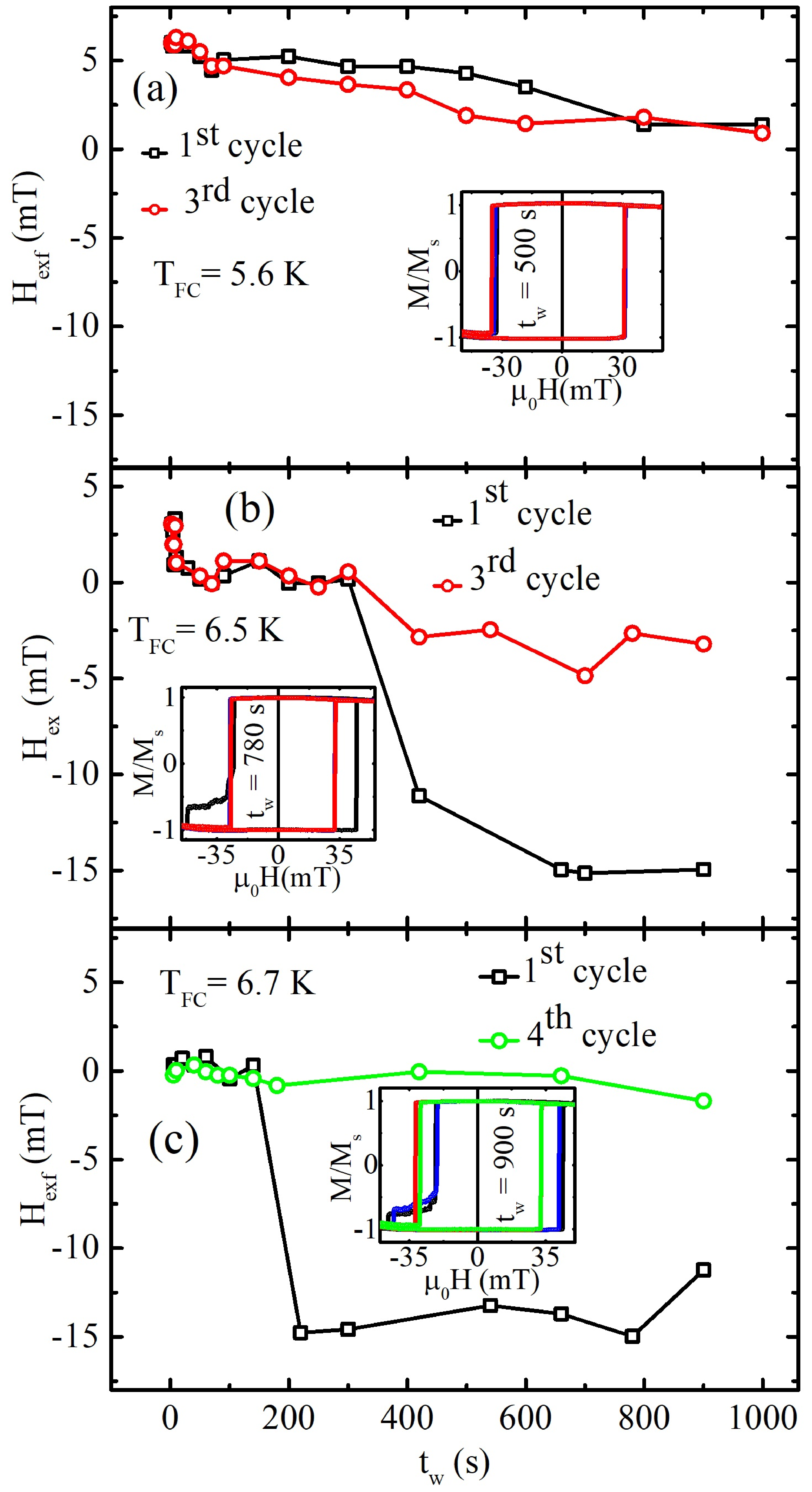}
	\caption{Variation of the exchange field $H_{\rm exf}$ with the waiting time $t_{\rm w}$, for the first M-H cycle (black) and the steady state (colored), for $T_{\rm FC}$ = 5.6, 6.5 and 7.0 K. The insets show the first three or four M-H loops at indicated $t_{\rm w}$ values.}
	\label{fig:YIGex6}
\end{figure}

The non-monotonic dependence of $\tau_0$ on $T_{\rm FC}$ in \mbox{Fig.} \ref{fig:YIGex4}(b) is somewhat unexpected. The peak in $\tau_0$ and jump in $H_{\rm exf}^{\rm b}$, both occurring near $T_{\rm FC}=6$ K, seem correlated. For a single energy barrier at play, $\tau_0$ should decrease monotonically with rising $T_{\rm FC}$ and become negligible when $T_{\rm FC}$ exceeds the blocking temperature. Spin-glasses exhibit a blocking temperature that depends on measurement time due to a wide distribution of energy-barriers as well as relaxation times. The $\tau_0$ behavior in \mbox{Fig.} \ref{fig:YIGex4}(b) may arise from a certain distribution of energy barriers due to disorder. With rising temperature, the lower barrier metastable spin configurations would first melt, enabling a change in $H_{\rm exf}$. The exact dependence of $\tau_0$ on $T_{\rm FC}$ will depend on the exact barrier distribution while $H_{\rm exf}^{\rm b}$ and $H_{\rm exf}^{\rm s}$ will depend on the average spin-order of the associated metastable states. Occurrence of two nearby peaks in energy barrier distribution can possibly lead to an intermediate peak in $\tau_0$ vs $T_{\rm FC}$.

\section{Training effect}
Training effects are common in exchange-bias systems \cite{hochstrat2002training, binek2004training} where the exchange bias is found to change with repeated field cycling. Figure \ref{fig:YIGex6}(a) shows the $t_{\rm w}$-dependent $H_{\rm exf}$ for the first and the third M-H loops for $T_{\rm FC}=5.6$ K, those loops being shown in the inset for a particular $t_{\rm w}$ value. The repeated M-H loops for several different $t_{\rm w}$ and $T_{\rm FC}$ can be found in the suppl-info \cite{suppl-info}. Here, up to $t_{\rm w}\sim80$ s, there is no training effect, but beyond this, $H_{\rm exf}$ reduces by 1 to 2 mT at the third loop as compared to the first. For $T_{\rm FC}=$ 6.0 and 6.7 K, see \mbox{Fig.} \ref{fig:YIGex6}(b,c), the training effect is much more drastic. Again, for up to a certain $t_{\rm w}$ there is no training effect, but after this a large difference appears between the first and the steady state $H_{\rm exf}$. The first cycle for large $t_{\rm w}$ shows $H_{\rm exf}$ close to -15 mT but in the steady state $H_{\rm exf}$ increases to become close to zero. Some of the metastabilities are visible in the intermediate M-H loops which show a small jump in magnetic moment between $\mu_0H=-45$ to -50 mT, see \mbox{Fig.} \ref{fig:YIGex6}(c) inset.

In every case, the steady state is achieved in 3 to 5 cycles. For $T_{\rm FC}$ above 7 K or below 5.5 K, no measurable training effect was seen. These training effects are much more drastic then those in conventional FM/AFM systems \cite{nogues1999exchange} and may indicate metastability of some of the interface-layer's spin-states. The interface layer, sandwiched between spin-ordered YIG and a paramagnetic GGG, faces two extreme, spin-ordered and paramagnetic, respectively, boundary conditions. The former promotes order and the latter, a dynamic disorder. The training effects allude to an affinity to the $H_{\rm exf}\approx0$ state, which could be the maximally spin-disordered intrinsic state if there were no exchange-field of YIG.

\section{Conclusion}
In conclusion, our findings indicate that RF sputtered films of YIG on GGG with post annealing at 750$^\circ$C exhibit a nanometer-thin spin-glass-like interface layer due to Fe diffusion into GGG. The spin-disordered state amounting to near-zero $H_{\rm exf}$ is favored although any state with $H_{\rm exf}$ ranging between 12.7 and -15 mT can be obtained by a suitable field-cooling protocol that leads to an arrest of the interface-layer's spins into a partially ordered state. While a complete understanding of the YIG/GGG interface needs more work, a tunable exchange bias in the lowest damping magnetic system YIG can be of significant application potential in memory devices as well as analog reservoir computing.

\section{ACKNOWLEDGEMENTS}
We would like to thank Robert Stamps for his critical views on this work. AKG acknowledges funding from the SERB-DST of the Government of India and from IIT Kanpur.

\clearpage
\setcounter{figure}{0}
\makeatletter
\renewcommand{\thefigure}{S\@arabic\c@figure}
\makeatother
\section{Supplemental Information}
There are six subsections in this suppl-info: 1) Elemental maps of Y, Gd and Ga, 2) M-H loop anisotropy, 3) Field cooling flow-chart, 4) $t{\rm w}$ dependent M-H loops, 5) $T_{\rm FC}$ dependent $H_{\rm exf}-t_{\rm w}$ data and, 6) $t{\rm w}$ and $T_{\rm FC}$ dependent repeated M-H loops.

\subsection{1) Elemental maps of Y, Gd and Ga}
\begin{figure}[hbt]
	\centering
	\includegraphics[width=3.4in]{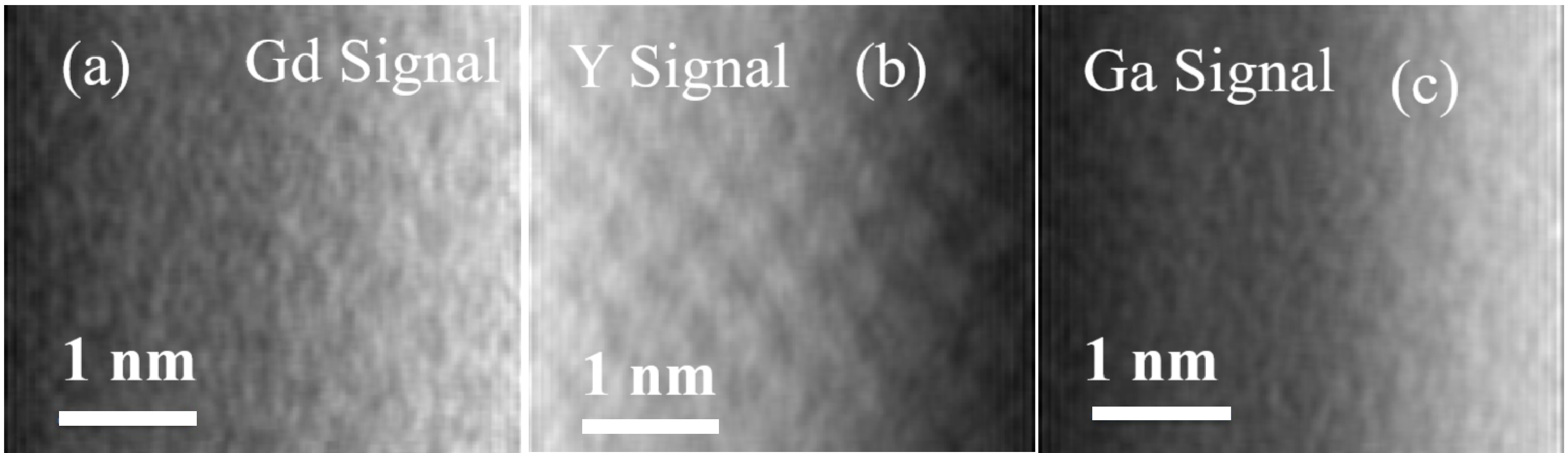}
	\caption{The maps of Y, Gd and Ga atoms acquired simultaneously with the image of Fig. \ref{fig:sfig1}(b) of the main paper. These images were used to find the line profiles presented in Fig. \ref{fig:YIGex2}(d) of the main paper.}
	\label{fig:sfig1}
\end{figure}

Figure \ref{fig:sfig1} shows the elemental maps of Y, Gd and Ga near the YIG/GGG interface. Note that no noticeable variation in oxygen signal was found in the interface region.
\begin{figure}[ht]
	\centering
	\includegraphics[width=3.0in]{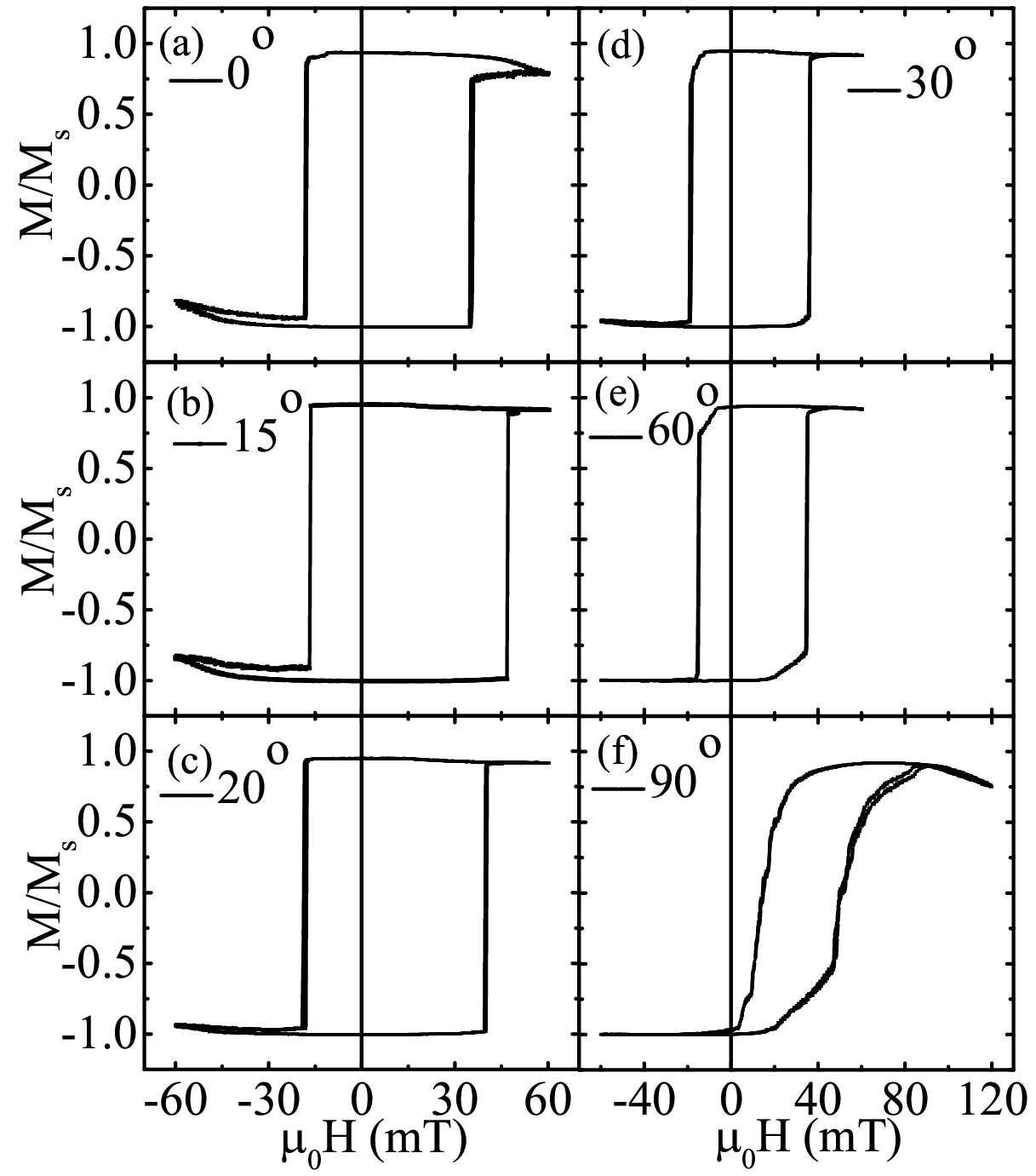}
	\caption{M-H loops for different in-plane magnetic field directions (0$^\circ$, 15$^\circ$, 20$^\circ$, 30$^\circ$, 60$^\circ$, 90$^\circ$) relative to the $\mu$-SQUID.}
	\label{fig:sfig2}
\end{figure}
\subsection{2) M-H loop anisotropy}
Figure \ref{fig:sfig2} shows the M-H loops for applied field along different directions, within the SQUID plane. The field angles are measured with respect to the axis of a magnet. From this the easy axis direction, exhibiting largest coercivity, was identified as 15$^\circ$. All subsequent data were collected with the applied field aligned along this 15$^\circ$ direction.
\subsection{3) Field cooling flow-chart}
\begin{figure}[ht]
	\centering
	\includegraphics[width=3.2in]{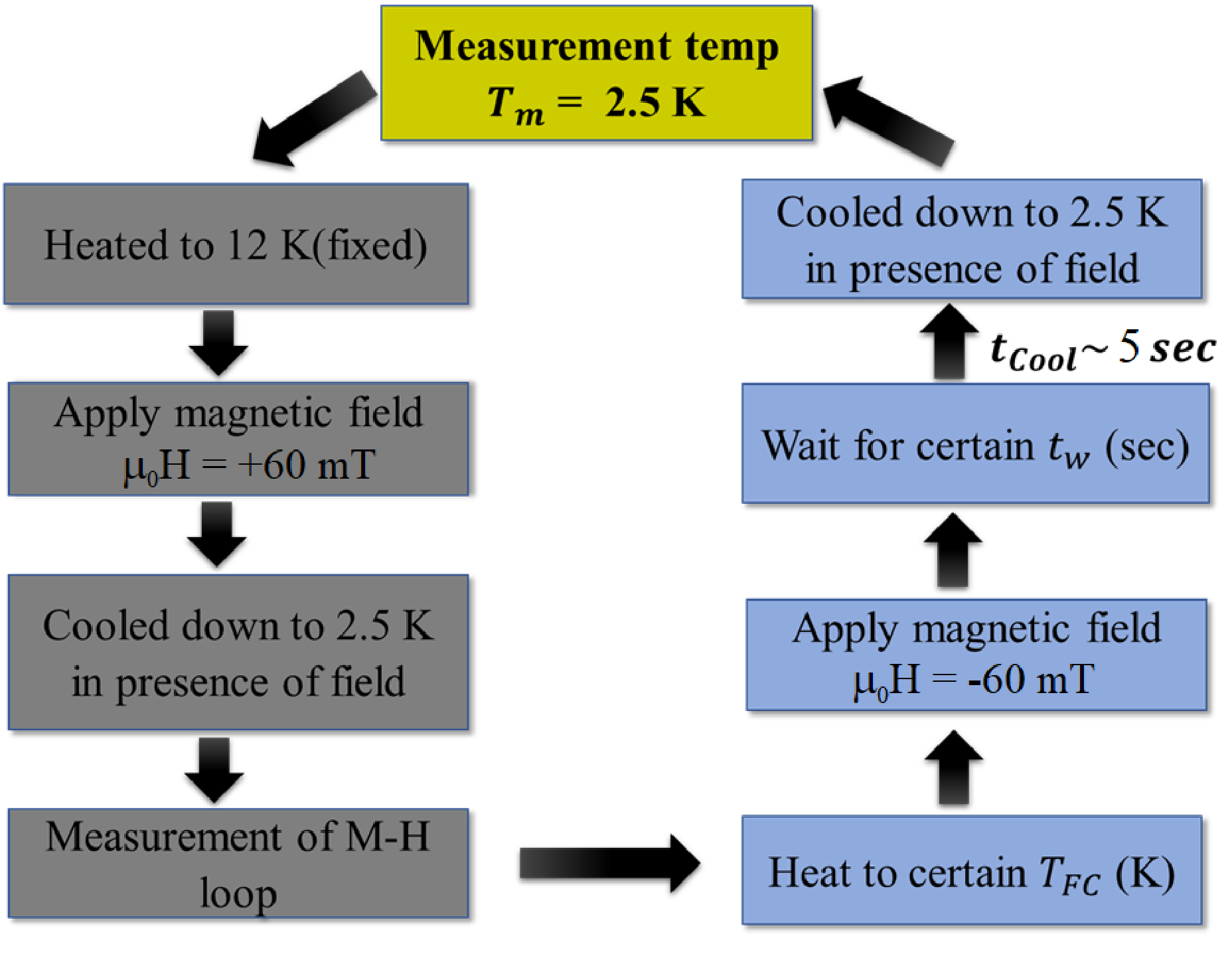}
	\caption{
Schematic of flow chart illustrating the field-cooling protocols, highlighting the measurement temperature of 2.5 K and the application of field cooling from various elevated temperatures $T_{FC}$. Here, the grey panels represent the reset while the blue ones represent the field cooling.}
	\label{fig:sfig3}
\end{figure}

Figure \ref{fig:sfig3} depicts the flow chart of one full cycle of the field cooling protocol including the reset at 12 K temperature and +60 mT field (grey panels). This protocol begins and ends at 2.5 K temperature, at which all the magnetic measurements were performed. Such a protocol with field cooling temperature $T_{\rm FC}<7$ K leads to a positive exchange field $H_{\rm exf}$. The same cycle but with opposite applied fields, both for reset and field cooling, was used for achieving a negative $H_{\rm exf}$ with $T_{\rm FC}<7$ K.

\subsection{4) $t_{\rm w}$ dependent M-H loops}
\begin{figure}[ht]
	\centering
	\includegraphics[width=3.4in]{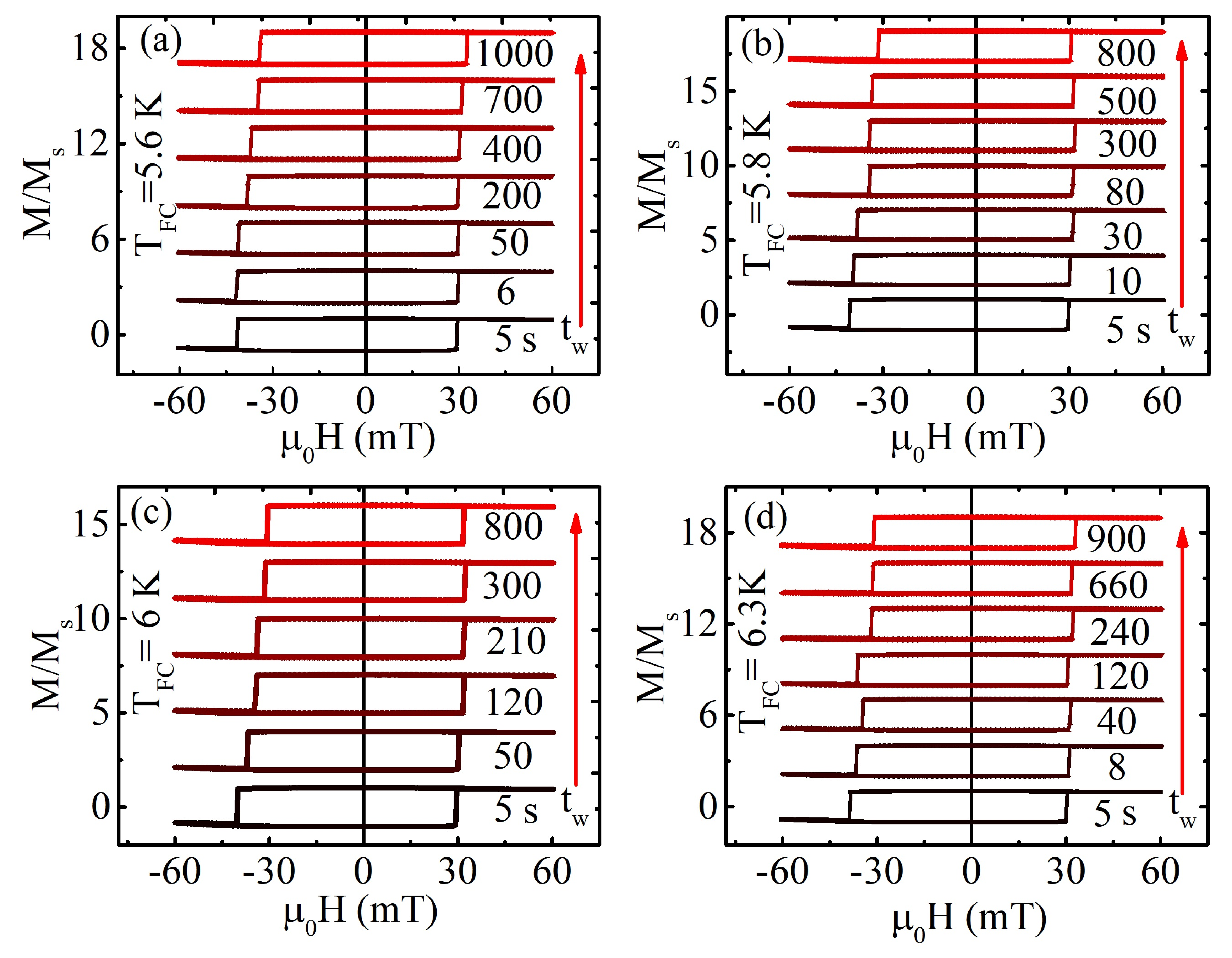}
	\caption{Evolution of M-H loops with $t_w$ for different $T_{FC}$ values as 5.6 (a), 5.8 (b), 6 (c) and 6.3 K (d).}
	\label{fig:sfig5}
\end{figure}

Figure \ref{fig:sfig5} shows the variation of steady state, \emph{i.e.} 3rd or 4th, M-H loops with $t_w$ for different $T_{FC}$ values. These loops have been used to deduce $H_{\rm exf}$ in the plots of Fig. \ref{fig:YIGex4}(a) of the main paper.

\subsection{5) $T_{\rm FC}$ dependent $H_{\rm exf}-t_{\rm w}$ data}
\begin{figure}[ht]
	\centering
	\includegraphics[width=3.4in]{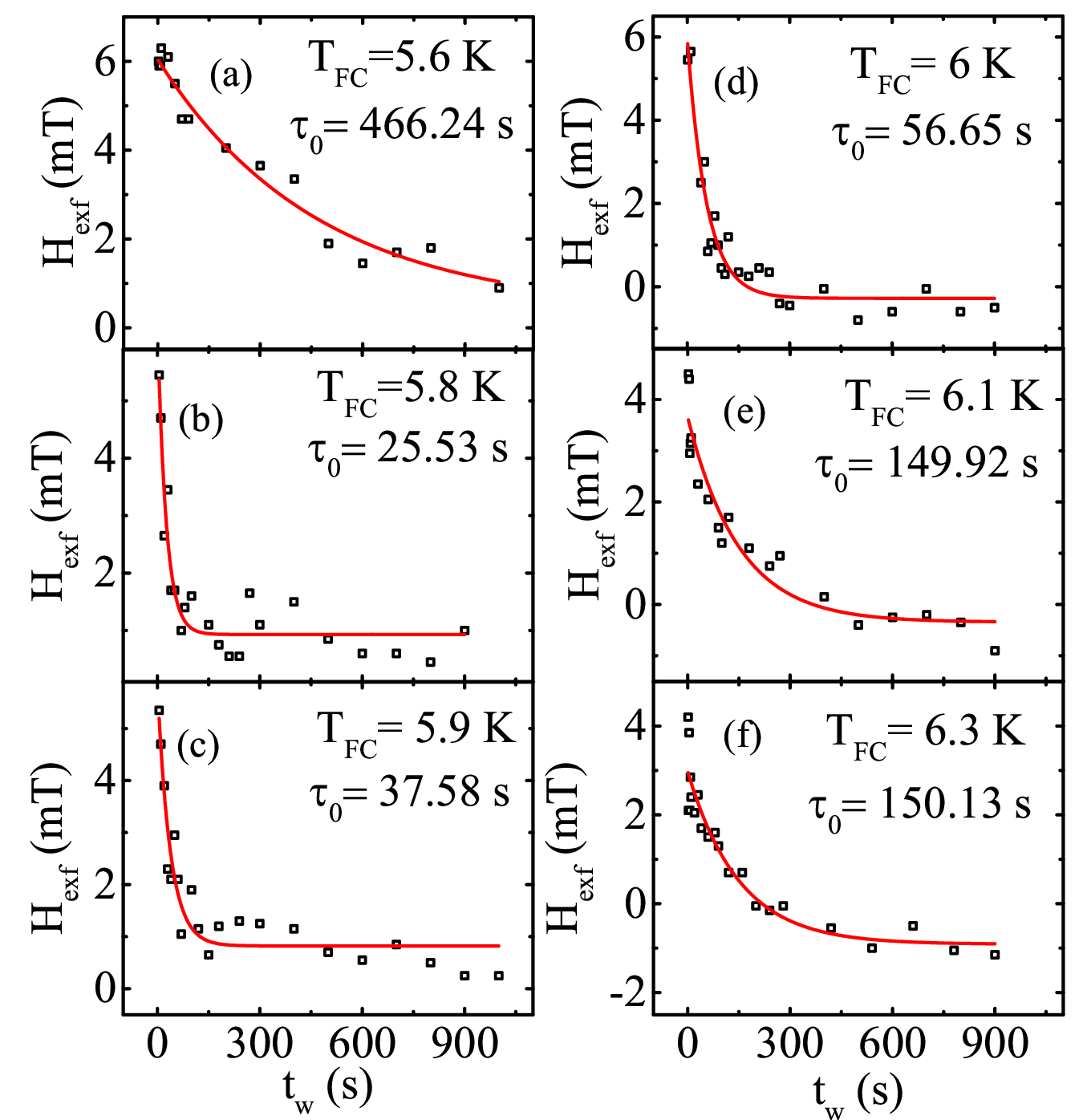}
	\caption{Variation of the exchange field $H_{\rm exf}$ with the waiting time $t_w$ for six different field-cooling temperature $T_{FC}$ values. These have been deduced from the M-H loops like those of Fig. \ref{fig:sfig5}.}
	\label{fig:sfig6}
\end{figure}

Figure \ref{fig:sfig6} shows the variation of the exchange field $H_{\rm exf}$ with the waiting time $t_w$ for six different $T_{FC}$ together with the fits to Eq. \ref{eq:FSS}. The fitting parameters have been summarized in Fig. \ref{fig:YIGex4}(b,c) of the main paper.
\subsection{6) $t{\rm w}$ and $T_{\rm FC}$ dependent repeated M-H loops}
\begin{figure}[hbt!]
	\centering
	\includegraphics[width=3.2in]{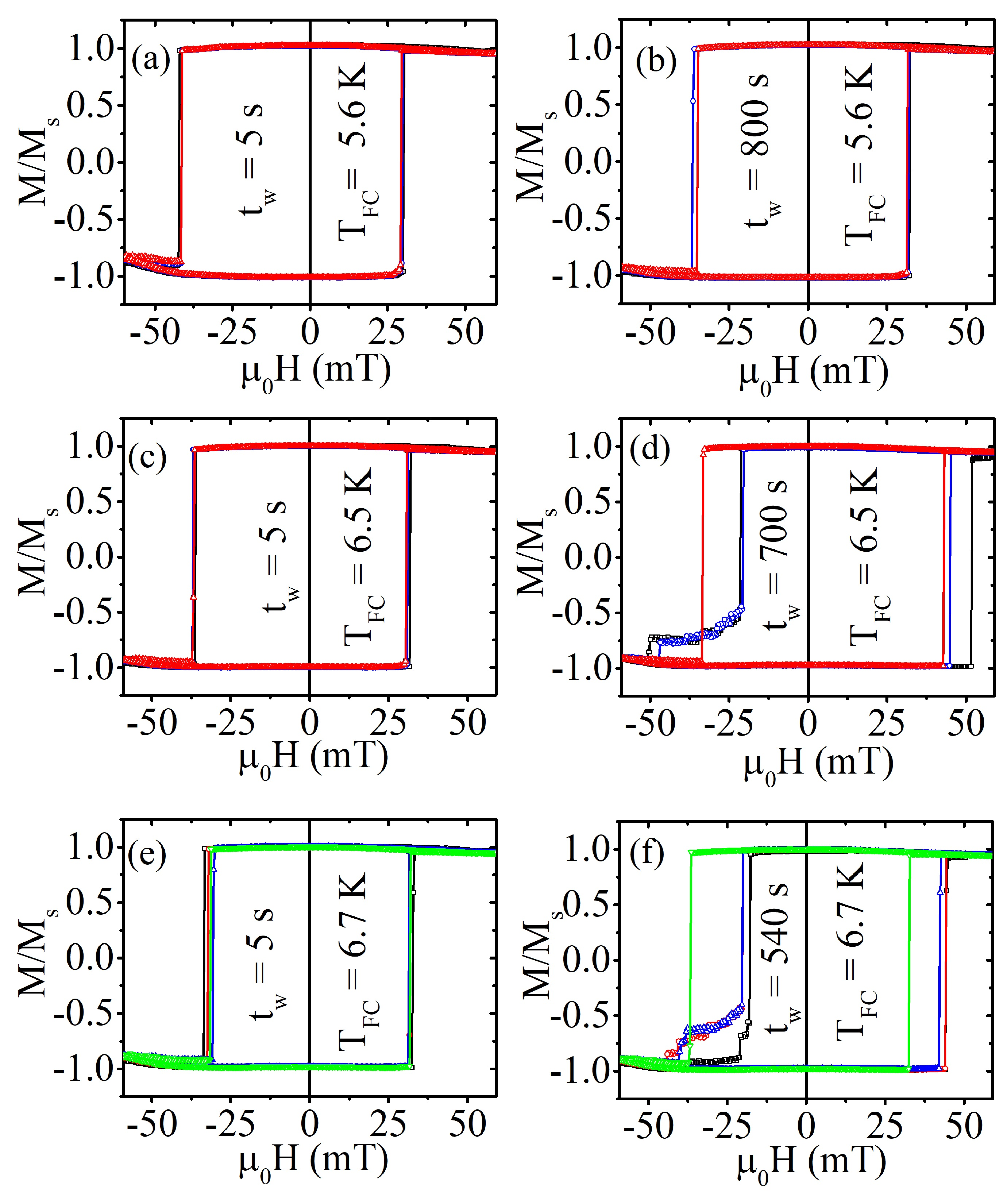}
	\caption{Repeated M-H loops for different values of $T_{\rm FC}$ and $t_{\rm w}$.}
	\label{fig:sfig7}
\end{figure}

Figure \ref{fig:sfig7} shows repeated M-H loops for different values of $T_{\rm FC}$ and $t_{\rm w}$ reflecting the training effect. The $H_{\rm exf}$ values deduced from these loops have been plotted in Fig. \ref{fig:YIGex6} of the main paper.


\begin{thebibliography}:
\bibitem{wu2010reversible}S.~M.~Wu, S.~A.~Cybart, P.~Yu, M.~D.~Rossell, J.~X.~Zhang, R.~Ramesh, and R.~C.~Dynes,  ``Reversible electric control of exchange bias in a multiferroic field-effect device", Nat. Mater. \textbf{9}, 756 (2010).
\bibitem{vzutic2004spintronics}I.~Žutić, J.~Fabian, and S.~Das~Sarma,  ``Spintronics: Fundamentals and applications", Rev. Mod. Phys. \textbf{76}, 323 (2004).
\bibitem{barman20212021}A.~Barman, G.~Gubbiotti, S.~Ladak, A.~O.~Adeyeye, M.~Krawczyk, J.~Gräfe, C.~Adelmann, S.~Cotofana, A.~Naeemi, V.~I.~Vasyuchka \textit{et al.}, ``The 2021 magnonics roadmap", J. Phys.: Condens. Matter \textbf{33}, 413001 (2021).
\bibitem{meiklejohn1956new} W. H. Meiklejohn and C. P. Bean,``New magnetic anisotropy", Phys. Rev. {\bf 102}, 1413 (1956)
\bibitem{kiwi2001exchange} M. Kiwi, ``Exchange bias theory", J. Mag. Magn. Mater. {\bf 234}, 584 (2001).
\bibitem{stamps2000mechanisms} R. L. Stamps, ``Mechanisms for exchange bias", J. Phys. D: Appl. Phys. {\bf 33}, R247 (2000).
\bibitem{malozemoff1987random} A. P. Malozemoff, ``Random-field model of exchange anisotropy at rough ferromagnetic-antiferromagnetic interfaces", Phys. Rev. B {\bf 35}, 3679 (1987).
\bibitem{nogues1999exchange} J. Nogu{\'e}s and I. K. Schuller, ``Exchange bias", J. Mag. Magn. Mater. {\bf 192}, 203 (1999).
\bibitem{NRC-FE1} K. Toprasertpong, E. Nako, Z. Wang, et al. ``Reservoir computing on a silicon platform with a ferroelectric field-effect transistor", Nat Commun Eng {\bf 1}, 21 (2022).
\bibitem{yuan2022quantum} H.~Y.~Yuan, Y.~Cao, A.~Kamra, R.~A.~Duine, and P.~Yan,  ``Quantum magnonics: When magnon spintronics meets quantum information science", Phys. Rep. \textbf{965}, 1 (2022).
\bibitem{lachance2019hybrid} D.~Lachance-Quirion, Y.~Tabuchi, A.~Gloppe, K.~Usami, and Y.~Nakamura,  ``Hybrid quantum systems based on magnonics", Appl. Phys. Express \textbf{12}, 070101 (2019).
\bibitem{cherepanov1993saga} V. Cherepanov, I. Kolokolov, and V. L'vov, ``The saga of YIG: Spectra, thermodynamics, interaction and relaxation of magnons in a complex magnet", Phys. Rep. {\bf 229}, 81 (1993).
\bibitem{serga2010yig} A. Serga, A. Chumak, and B. Hillebrands, ``YIG magnonics", J. Phys. D: Appl. Phys. {\bf 43}, 264002 (2010).
\bibitem{hauser2016yttrium} C. Hauser, T. Richter, N. Homonnay, C. Eisenschmidt, M. Qaid, H. Deniz, D. Hesse, M. Sawicki, S. G. Ebbinghaus, and G. Schmidt, ``Yttrium iron garnet thin films with very low damping obtained by recrystallization of amorphous material", Sci. Rep. {\bf 6}, 20827 (2016).
\bibitem{popova2003exchange} E. Popova, N. Keller, F. Jomard, L. Thomas, M.-C. Brianso, F. Gendron, M. Guyot, and M. Tessier, ``Exchange coupling in ultrathin epitaxial yttrium iron garnet films", Eur. Phys. J. B {\bf 31}, 69 (2003).
\bibitem{roos2022magnetization} M. Roos, P. Quarterman, J. Ding, M. Wu, B. Kirby, and B. Zink, ``Magnetization and antiferromagnetic coupling of the interface between a 20 nm Y$_3$Fe$_5$O$_{12}$ film and Gd$_3$Ga$_5$O$_{12}$ substrate", Phys. Rev. Mater. {\bf 6}, 034401 (2022).
\bibitem{kumar2021positive} R. Kumar, S. Sarangi, D. Samal, and Z. Hossain, ``Positive exchange bias and inverted hysteresis loop in Y$_3$Fe$_5$O$_{12}$/Gd$_3$Ga$_5$O$_{12}$", Phys. Rev. B {\bf 103}, 064421 (2021).
\bibitem{mitra2017interfacial} A. Mitra, O. Cespedes, Q. Ramasse, M. Ali, S. Marmion, M. Ward, R. Brydson, C. Kinane, J. Cooper, S. Langridge, \emph{et. al.}, ``Interfacial origin of the magnetisation suppression of thin film yttrium iron garnet", Sci. Rep. \textbf{7}, 11774 (2017).
\bibitem{suturin2018role} S. Suturin, A. Korovin, V. Bursian, L. Lutsev, V. Bourobina, N. Yakovlev, M. Montecchi, L. Pasquali, V. Ukleev, A. Vorobiev, et al., ``Role of gallium diffusion in the formation of a magnetically dead layer at the Y$_3$Fe$_5$O$_{12}$/Gd$_3$Ga$_5$O$_{12}$ epitaxial interface", Phys. Rev. Mater. {\bf 2}, 104404 (2018).
\bibitem{gomez2018synthetic} J. M. Gomez-Perez, S. V\'{e}lez, L. McKenzie-Sell, M. Amado, J. Herrero-Mart\'{i}n, J. L\'{o}pez-L\'{o}pez, S. Blanco-Canosa, L. E. Hueso, A. Chuvilin, J. W. Robinson, and F. Casanova, ``Synthetic antiferromagnetic coupling between ultrathin insulating garnets", Phys. Rev. Appl. {\bf 10}, 044046 (2018).
\bibitem{cannella1972magnetic} V. Cannella and J. A. Mydosh, ``Magnetic ordering in gold-iron alloys", Phys. Rev. B {\bf 6}, 4220 (1972).
\bibitem{ali2007exchange} M. Ali, P. Adie, C. H. Marrows, D. Greig, B. J. Hickey, and R. L. Stamps, ``Exchange bias using a spin glass", Nat. Mater. { \bf 6}, 70 (2007).
\bibitem{fisher1988nonequilibrium} D. S. Fisher and D. A. Huse, ``Nonequilibrium dynamics of spin glasses", Phys. Rev. B {\bf 38}, 373 (1988).
\bibitem{dotsenko1993physics} V. S. Dotsenko, ``Physics of the spin-glass state", Physics-Uspekhi {\bf 36}, 455 (1993).
\bibitem{jonason1998memory} K. Jonason, E. Vincent, J. Hammann, J. Bouchaud, and P. Nordblad, ``Memory and chaos effects in spin glasses", Phys. Rev. Lett. {\bf 81}, 3243 (1998).
\bibitem{ferreiro2014frustration}D. U. Ferreiro, E. A. Komives, and P. G. Wolynes, ``Frustration in biomolecules", Q. Rev. Biophys. {\bf 47}, 285 (2014).
\bibitem{moessner2006geometrical}R.~Moessner and A.~P.~Ramirez,  ``Geometrical frustration", Phys. Today \textbf{59}, 24 (2006).
\bibitem{paul2020probing} S. Paul, G. Kotagiri, R. Ganguly, H. Parashari, H. Courtois, C. B. Winkelmann, and A. K. Gupta, ``Probing magnetism of individual nano-structures using Nb-SQUIDs in hysteresis free mode", J. Mag. Mag. Mater. {\bf 503}, 166625 (2020).
\bibitem{suppl-info} See Supplementary Information for other elements' EDX images, M-H loop anisotropy, flow-chart of the field-cooling protocol, $t{\rm w}$ dependent M-H loops, more extensive $T_{\rm FC}$ dependent $H_{\rm exf}-t_{\rm w}$ data and $t_{\rm w}$ and $T_{\rm FC}$ dependence of repeated M-H loops for training effect.
\bibitem{medwal2021facet} R. Medwal, A. Deka, J. V. Vas, M. Duchamp, H. Asada, S. Gupta, Y. Fukuma, and R. S. Rawat, ``Facet controlled anisotropic magnons in Y$_3$Fe$_5$O$_{12}$ thin films", Appl. Phys. Lett. {\bf 119} (2021).
    \bibitem{musa2017structural}M.~A.~Musa, R.~S.~Azis, N.~H.~Osman, J.~Hassan, and T.~Zangina,
``Structural and magnetic properties of yttrium iron garnet (YIG) and yttrium aluminum iron garnet (YAlG) nanoferrite via sol-gel synthesis",  Results Phys. \textbf{7}, 1135 (2017).
\bibitem{petrenko1999magnetic} O. Petrenko, D. M. Paul, C. Ritter, T. Zeiske, and M. Yethiraj, ``Magnetic frustration and order in gadolinium gallium garnet", Physica B {\bf 266}, 41 (1999).
\bibitem{schiffer1995frustration}P.~Schiffer, A.~P.~Ramirez, D.~A.~Huse, P.~L.~Gammel, U.~Yaron, D.~J.~Bishop, and A.~J.~Valentino,  ``Frustration induced spin freezing in a site-ordered magnet: Gadolinium gallium garnet", Phys. Rev. Lett. \textbf{74}, 2379 (1995).
\bibitem{gorbatov2021magnetic}O.~I.~Gorbatov, G.~Johansson, A.~Jakobsson, S.~Mankovsky, H.~Ebert, I.~Di~Marco, J.~Minár, and C.~Etz,  ``Magnetic exchange interactions in yttrium iron garnet: A fully relativistic first-principles investigation", Phys. Rev. B \textbf{104}, 174401 (2021).
\bibitem{note-diff-runs} In different experimental runs, starting from room temperature and over a period of several months, the value of $T_{\rm m}$, the values of the positive and negative extremes of $H_{\rm exf}$ were found to be slightly different. An estimate of these spreads from various runs is: $\Delta T_{\rm m}\approx 0.5$ K and $\Delta H^{\rm max}_{\rm exf}\approx 2$ mT. However, the spreads were much smaller in a particular run. The difference in the magnitudes of the two extremes of $H_{\rm exf}$ amounts to an asymmetry that varies between runs. We believe that this arises from the magnetic subsystems that get blocked at temperatures much higher than the used reset temperature of 12 K. This is beyond the scope of this work. Note also that the data presented in \mbox{Fig.} \ref{fig:YIGex3}(d) for two opposite reset fields are from two different runs.
\bibitem{vincent2022spin}E.~Vincent,  ``Spin glass experiments,''  arXiv:2208.00981 (2022).
\bibitem{hochstrat2002training}A.~Hochstrat, Ch.~Binek, and W.~Kleemann,``Training of the exchange-bias effect in NiO-Fe heterostructures", Phys. Rev. B \textbf{66}, 092409 (2002).
\bibitem{binek2004training}Ch.~Binek,``Training of the exchange-bias effect: A simple analytic approach", Phys. Rev. B \textbf{70}, 014421 (2004).
\bibitem{binder1986spin}K.~Binder and A.~P.~Young,``Spin glasses: Experimental facts, theoretical concepts, and open questions", Rev. Mod. Phys. \textbf{58}, 801 (1986).
\bibitem{anderson1988spin}P.~W.~Anderson,``Spin glass I: A scaling law rescued", Phys. Today \textbf{41}, 9 (1988).
\end{thebibliography}
\end{document}